\documentclass[11pt]{article}

\usepackage{amsfonts}
\usepackage{amsmath}
\usepackage{amssymb}
\usepackage{amsthm}

\usepackage[cp1251]{inputenc}
\usepackage[russian]{babel}

\usepackage{graphicx}

\theoremstyle{plain}
\newtheorem{theorem}{Теорема}

\newtheorem{definition}[theorem]{Определение}

\newtheorem{corollary}[theorem]{Следствие}

\theoremstyle{definition}
\newtheorem{example}[theorem]{Пример}
\newtheorem{remark}[theorem]{Замечание}

\numberwithin{equation}{section}
\numberwithin{theorem}{section}

\DeclareMathOperator{\Int}{int}

\DeclareMathOperator{\bd}{bd}

\DeclareMathOperator{\diag}{diag}
\newcommand{\D}{\,\mathrm{d}}

\newcommand{\bs}[1]{{\boldsymbol{#1}}}

\newcommand{\pdt}{\partial_t}
\newcommand{\R}{\mathbf{R}}

\newcommand{\DP}[2]{\langle#1,#2\rangle}

\begin{document}
%The heading
\title{{Предельное поведение пространственно распределенных репликаторных систем}}

\author{А. С. Братусь$^{1,2}$, А. С. Новожилов$^{3}$, В. П. Посвянский$^{1}$ \\[3mm]
\textit{\normalsize $^\textrm{\emph{1}}$Прикладная Математика--1, Московский Государственный Университет }\\[-1mm]
\textit{\normalsize Путей Сообщения, Москва 127994, Россия}\\[2mm]
\textit{\normalsize $^\textrm{\emph{2}}$Факультет Вычислительной Математики и Кибернетики,}\\[-1mm]
\textit{\normalsize Московский Государственный Университет им. М.В. Ломоносова,}\\[-1mm]
\textit{\normalsize Москва 119992, Россия}\\[2mm]
\textit{\normalsize $^\textrm{\emph{3}}$Отделение Математики, Государственный Университет Северной Дакоты,}\\[-1mm]
\textit{\normalsize Фарго, Северная Дакота, 58108, США}}

\date{}

\maketitle

%Abstract goes here
\begin{abstract}
Исследуется вопрос биологической стабильности распределенной репликаторной системы типа реакция--диффузия. Найдены достаточные условия биологической стабильности, и доказано, что учет пространственной неоднородности влечет биологическую стабильность системы в некоторых случаях, когда нераспределенная (локальная) система не является биологически стабильной. Аналитические результаты иллюстрируются численными примерами.

\paragraph{\small Ключевые слова:}

\end{abstract}
\section{Мотивация} Цель данной работы --- представить достаточные условия \textit{биологической стабильности} для некоторого класса уравнений в частных производных, заданных на интегральном симплексе.

Дифференциальные уравнения --- один из наиболее распространенных математических аппаратов, использующихся в математической биологии для описания тех или иных биологических систем (см., например, книгу \cite{dinsys2010} и ссылки в ней). Например, одна из самых хорошо изученных систем обыкновенных дифференциальных уравнений (ОДУ) в биоматематике имеет вид
\begin{equation}\label{eq0:1}
\dot w_i=w_i\bigl(f_i(\bs w)-f^l(\bs w)\bigr),\quad i=1,\ldots,n,
\end{equation}
и называется \textit{репликаторным уравнением} \cite{hofbauer1998ega,hofbauer2003egd}. Здесь $\bs w=\bigl(w_1(t),\ldots,w_n(t)\bigr)$ --- вектор переменных, которые мы будем называть макромолекулами; таким образом, $w_i$ --- это относительная концентрация $i$-го типа макромолекул. Макромолекулы взаимодействуют друг с другом, и эти взаимодействия задают \textit{приспособленности} $f_i(\bs w)$ для каждого $i$. Выражение $f^l(\bs w)$ не зависит от индекса $i$ и необходимо, чтобы $\sum_{i=1}^nw_i=\text{const}$ для любого момента времени $t$. Отметим, что часто $f_i(\bs w)=\sum_{j=1}^n a_{ij}w_j$, и взаимодействия макромолекул задаются \textit{матрицей взаимодействий} $\bs A=(a_{ij})_{n\times n}$.

Наряду с \eqref{eq0:1} также часто рассматриваются так называемые системы (популяционной динамики) Колмогорова:
\begin{equation}\label{eq0:2}
    \dot W_i=W_if_i(\bs W),\quad i=1,\ldots,n,
\end{equation}
где $\bs W=\bigl(W_1(t),\ldots,W_n(t)\bigr)$ --- вектор абсолютных численностей взаимодействующих популяций, а $f_i(\bs W)$ --- снова приспособленности (интенсивность рождаемости минус интенсивность смертности). Формально, системы \eqref{eq0:1} и \eqref{eq0:2} эквивалентны, если предположить, что все $f_i$ --- однородные функции одной степени, и переход от \eqref{eq0:2} к \eqref{eq0:1} осуществляется рассмотрением уравнений для относительных численностей $w_i=W_i/\sum_{j=1}^n W_j$. Разумеется, необходимо также предположить, что $\sum_{j=1}^n W_j$ не стремится к нулю. Отметим, что $f^l(\bs w)=\sum_{i=1}^nw_if_i(\bs w)$.

Одно из центральных понятий теории дифференциальных уравнений --- понятие устойчивости, для которого существует множество (неэквивалентных) математических определений, самое распространенное из которых --- устойчивость по Ляпунову. Устойчивость по Ляпунову, однако, часто слишком ограничительна для анализа математических моделей вида \eqref{eq0:1} или \eqref{eq0:2}, так как с биологической точки зрения устойчивость системы наиболее часто означает ее неограниченное существование как целого, а не выход на какой-то один определенный динамический режим \cite{lewontin1968meaning}. Таким образом, необходимо точное понятие устойчивости для моделей типа \eqref{eq0:1} и \eqref{eq0:2}, которое бы формализовало тот факт, что при $t\in[0,\infty)$ ни одна из компонент векторов $\bs w$ и $\bs W$ не приближается опасно близко к нулю, и ни одна компонента вектора $\bs W$ не стремится к бесконечности. Такое определение --- обобщение математического понятия устойчивости по Лагранжу. В книге \cite{svireshev1978} данный тип устойчивости назван \textit{экологической стабильностью} (в \cite{dinsys2010} используется термин \textit{экологическая устойчивость}). В англоязычной литературе существует два (эквивалентных) определения данной устойчивости: \textit{перманентность} (``permanence'', введено в \cite{schuster1979dynamical}) и \textit{персистентность} (``persistence'', введено в \cite{freedman1977mathematical}, в современной литературе это понятие превратилось в ``weak persistence''). Поскольку системы вида \eqref{eq0:1} или \eqref{eq0:2} моделируют многие процессы за рамками того, что формально описывается \textit{экологическими} взаимодействиями, в данном тексте мы будем использовать термин \textit{биологическая стабильность} (точное определение см. в следующем пункте).

В области математического моделирования биологических систем существует следующее неформальное наблюдение: введение явной пространственной неоднородности в модель часто приводит к увеличению той области пространства параметров задачи, в которой система биологически стабильна \cite{dieckmann2000}. Более того, часто введение пространства в модель влечет биологическую стабильность полученной системы, не смотря на отсутствие биологической стабильности в нераспределенной системе.

Для задачи \eqref{eq0:2} учет пространственной неоднородности прост. Пусть теперь $\bs W=\bigl(W_1(\bs x,t),\ldots,W_n(\bs x,t)\bigr)$, где $\bs x\in\Omega\subseteq\R^m$, где $m$ может принимать значения 1, 2 или 3 в зависимости от размерности моделируемого пространства. Тогда динамика плотностей взаимодействующих популяций в области $\Omega$ описывается системой типа реакция--диффузия
\begin{equation}\label{eq0:3}
    \pdt W_i=W_if_i(\bs W)+d_i\Delta W_i,\quad i=1,\ldots,k,
\end{equation}
где $\Delta$ --- оператор Лапласа, а $d_i\geq 0$ --- коэффициенты диффузии. Необходимо также задать начальные и граничные условия.

Аналогичный подход невозможен для системы \eqref{eq0:1}, если не предполагать равенства всех $d_i$ \cite{hadeler1981dfs,novozhilov2012reaction,vickers1989spa,weinberger1991ssa}. В работах \cite{bratus2006ssc,bratus2009existence,bratus2011} рассматривался следующий распределенный аналог задачи \eqref{eq0:1}, в случае когда $f_i(\bs v)=\sum_{j=1}^na_{ij}v_j$:
\begin{equation}\label{eq0:4}
    \pdt v_i=v_i\bigl(f_i(\bs v)-f^{s}_1(\bs v)\bigr)+d_i\Delta v_i,\quad i=1,\ldots,n,
\end{equation}
где
$$
f_1^s(\bs v)=\int_\Omega\DP{\bs{Av}}{\bs v}\D \bs x,\quad \DP{\bs v}{\bs w}=\sum_{i=1}^n v_iw_i.
$$
Один из важных выводов для задачи \eqref{eq0:4}, который получен в процитированных выше работах, заключается в том, что таким образом введенная пространственная структура в репликаторное уравнение не приводит к б\'{о}льшей биологической стабильности системы. Для одних и тех же компонент матрицы взаимодействий $\bs A$ те характеристики систем \eqref{eq0:1} и \eqref{eq0:4}, которые определяют биологическую стабильность, совпадают.

Подход, приводящий к системе \eqref{eq0:4}, не является единственно возможным (другие подходы обсуждаются и сравниваются в \cite{novozhilov2012reaction}), и в \cite{novozhilov2013replicator} была предложена другая репликаторная система с пространственной структурой (см. ниже уравнение \eqref{eq1:1}). Численные эксперименты, представленные в \cite{novozhilov2013replicator}, показали, что для этой системы упомянутый выше неформальный вывод верен: введение пространственных переменных влечет в некоторых случаях биологическую стабильность системы, нестабильную в нераспределенном случае. В данной работе мы приводим доказательство данного факта и иллюстрируем его примерами.

\section{Постановка задачи}
Пусть $\Omega$ --- ограниченная область в $\R^m$, где $m$ равно 1, 2, или 3, с кусочно-гладкой границей $\Gamma$, $\bs A=(a_{ij})_{n\times n}$ --- заданная действительная матрица, $\bs v=\bs v(\bs x,t)=\bigl(v_1(\bs x,t),\ldots,\bs v_n(\bs x,t)\bigr)$ --- вектор-функция, $\bs x\in \Omega$, $t\geq 0$. Введем обозначения
\begin{align*}
\bigl(\bs A\bs v\bigr)_k&=\sum_{j=1}^n a_{kj}v_j(\bs x,t),\quad k=1,\ldots,n,\\
\DP{\bs{Av}}{\bs v}&=\sum_{j=1}^n\bigl(\bs{Av}\bigr)_jv_j(\bs x,t)=\sum_{j,k=1}^na_{kj}v_k(\bs x,t)v_j(\bs x,t).
\end{align*}

Рассмотрим начально-краевую задачу для следующей системы ($d_k>0,\,k=1,\ldots,n$ --- параметры)
\begin{equation}\label{eq1:1}
    \pdt v_k=v_k\Bigl(\bigl(\bs A\bs v\bigr)_k-f^s(\bs v)+d_k\Delta v_k\Bigr),\quad k=1,\ldots,n,
\end{equation}
с начальными и краевыми условиями
\begin{equation}\label{eq1:2}
    v_k(\bs x,0)=\varphi_k(\bs x),\quad \left.\frac{\partial v_k(\bs x,t)}{\partial\nu}\right|_{\bs x\in \Gamma}=0,\quad k=1,\ldots,n,
\end{equation}
где $\nu$ --- внешняя нормаль к $\Gamma$. В системе \eqref{eq1:1}
\begin{equation}\label{eq1:3}
    f^s(\bs v)=\int_{\Omega}\Bigl(\DP{\bs{Av}}{\bs v}-\sum_{k=1}^n d_k\|\nabla v_k\|^2\Bigr)\D \bs x.
\end{equation}
Отметим, что система \eqref{eq1:1}--\eqref{eq1:3} не является системой уравнений в частных производных в общепринятом смысле, так как $f^s(\bs v)$ является функционалом над решениями задачи \eqref{eq1:1}--\eqref{eq1:2}.

Из \eqref{eq1:1}--\eqref{eq1:3} следует, что
$$
\frac{\partial}{\partial t}\left(\sum_{k=1}^n\int_{\Omega} v_k(\bs x,t)\D \bs x\right)=0,
$$
что означает
\begin{equation}\label{eq1:4}
    \sum_{k=1}^n\int_{\Omega} v_k(\bs x,t)\D \bs x=\text{const}
\end{equation}
для любого $t\geq 0$, где константа может быть выбрана произвольно и, например, взята равной единице. То есть интегральный симплекс решений (см. ниже) задачи \eqref{eq1:1}--\eqref{eq1:4} инвариантен.

Задача \eqref{eq1:1}--\eqref{eq1:2} является распределенным репликаторным уравнением типа реакция--диффузия и описывает, например, популяционную динамику сообщества самовоспроизводящихся макромолекул. Тогда $v_k(\bs x,t)$ --- относительная плотность макромолекул $k$-го типа по отношению к суммарной численности всего сообщества макромолекул в области $\Omega$  в момент времени $t$. Функционал $f^s(\bs v)$ в такой интерпретации называется средней приспособленностью популяции, а выражение $\bigl(\bs A\bs v\bigr)_k$ --- приспособленностью $k$-го вида макромолекул в точке $\bs x\in\Omega$ в момент времени $t$.

Из физического смысла задачи следует, что решение системы \eqref{eq1:1}--\eqref{eq1:3} следует искать среди множества неотрицательных функций $v_k(\bs x,t)\geq 0,\,\bs x\in\Omega,\,t\geq 0,k=1,\ldots,n$.

Далее будем полагать, что функции $v_k(\bs x,t)\geq 0,\,\bs x\in\Omega,\,t\geq 0,k=1,\ldots,n$ являются гладкими по переменной $t$ и вместе со своими производными по переменной $t$ принадлежат пространству Соболева $W^{1,2}$, если $m=1$, и пространству $W^{2,2}$, если $m=2,3$, для каждого фиксированного значения $t>0$. Здесь $W^{s,2}$ --- пространство функций интегрируемых с квадратом в $\Omega$ вместе со своими (слабыми) производными до порядка $s$. Отметим, что из теорем вложения (например, \cite{evans_2010}) следует, что такие функции совпадают с непрерывними функциями почти всюду в $\Omega$.

Обозначим $\Omega_t=\Omega\times[0,\infty)$ и рассмотрим пространство функций $B(\Omega_t)$ с нормой
$$
\|z(\bs x,t)\|_{B(\Omega_t)}=\max_{t\geq 0}\left\{\|z(\bs x,t)\|_{W^{s,2}}+\|\pdt z(\bs x,t)\|_{W^{s,2}}\right\},\quad s=1,2.
$$
Через $S_n(\Omega_t)$ обозначим множество неотрицательных функций $\bs v(\bs x,t)=\bigl(v_1(\bs x,t),\ldots,v_n(\bs x,t)\bigr)$ таких, что $v_k(\bs x,t)\in B(\Omega_t)$  для всех $k$, и выполняется условие \eqref{eq1:4} с постоянной равной единице:
\begin{equation}\label{eq1:5}
    \sum_{k=1}^n\int_{\Omega} v_k(\bs x,t)\D \bs x=1.
\end{equation}
Множество $S_n(\Omega_t)$ --- интегральный симплекс в пространстве вектор-функций, каждая компонента которых принадлежит $B(\Omega_t)$.

Граничными элементами (обозначение $\bd S_n(\Omega_t)$) интегрального симплекса $S_n(\Omega_t)$ назовем множество вектор-функций $\bs v(\bs x,t)=\bigl(v_1(\bs x,t),\ldots,v_n(\bs x,t)\bigr)$ таких, что для непустого множества индексов $K_0\subset \{1,\ldots,n\}$
$$
\overline{v}_k(t)=\int_\Omega v_k(\bs x,t)\D \bs x=0,\quad k\in K_0,
$$
и $\overline{v}_k(t)>0$, $k\notin K_0$, $t\geq 0$. В силу инвариантности симплекса
\begin{equation}\label{eq1:6}
    \sum_{k\notin K_0}\overline{v}_k(t)=1.
\end{equation}
Соответственно, внутренними элементами интегрального симплекса $S_n(\Omega_t)$ (обозначение $\Int S_n(\Omega_t)$) будем называть такие вектор-функции $\bs v(\bs x,t)\in S_n(\Omega_t)$, для компонент которых
$$
\overline{v}_k(t)=\int_\Omega v_k(\bs x,t)\D \bs x>0,\quad k=1,\ldots,n,\quad t\geq 0.
$$
Далее, не умаляя общности, считаем, что мера области $\Omega$ равна единице, т.е., $|\Omega|=1$.
\begin{remark}Так как неотрицательные функции $v_k(\bs x,t)\in W^{s,2},\,k=1,\ldots,n$ для $s=1$ или $s=2$ при каждом $t\geq 0$, то из теоремы вложения следует, что они совпадают почти всюду с непрерывными функциями. Следовательно, учитывая неотрицательность функций, можно заключить, что если среднее интегральное значение $\overline{v}_k(t)=0$, то $v_k(\bs x,t)=0$ почти всюду в $\Omega$. Таким образом, множество $\bd S_n(\Omega_t)$ состоит из вектор-функций, для которых
$$
v_k(\bs x,t)=0,\quad k\in K_0,
$$
и выполняется равенство \eqref{eq1:6}.
\end{remark}

Решение задачи \eqref{eq1:1}--\eqref{eq1:3} будем понимать в слабом смысле. Вектор-функция $\bs v(\bs x,t)\in S_n(\Omega_t)$ является слабым решением, если выполняется интегральное тождество
\begin{align*}
\int_0^\infty\int_\Omega \pdt v_k(\bs x,t)\eta(\bs x,t)\D \bs x\D t&=\int_0^\infty\int_\Omega v_k(\bs x,t)\Bigl(\bigl(\bs{Av}\bigr)_k-f^s(\bs v)\Bigr)\eta(\bs x,t)\D \bs x\D t\\
&-d_k\int_0^\infty\int_\Omega \DP{\nabla v_k(\bs x,t)}{\nabla \eta(\bs x,t)}\D \bs x\D t,
\end{align*}
для любой функции $\eta(\bs x,t)$, которая при $\bs x\in\Omega$ дифференцируема по $t$ и имеет компактный носитель при любом фиксированном  $t\in[0,\infty)$, и при любом $t\geq 0$ принадлежит пространству $W^{s,2}(\Omega)$ для $s=1$ или $s=2$.

Наряду с системой \eqref{eq1:1}--\eqref{eq1:3} рассмотрим систему обыкновенных дифференциальных уравнений, которая формально получается из исходной задачи в пределе $d_k\to 0$:
\begin{equation}\label{eq1:7}
    \dot w_k=w_k\Bigl(\bigl(\bs{Aw}\bigr)_k-f^l(\bs w)\Bigr),\quad k=1,\ldots n,
\end{equation}
с начальными условиями
$$
w_k(0)=w_k^0,\quad k=1,\ldots,n.
$$
Здесь
$$
f^l(\bs w)=\DP{\bs{Aw}}{\bs w}=\sum_{i,j=1}^n a_{ij}w_iw_j.
$$
Система \eqref{eq1:7} рассматривается на множестве неотрицательных гладких вектор-функций $\bs w=\bs w(t)=\bigl(w_1(t),\ldots,w_n(t)\bigr)$, принадлежащих для каждого момента времени симплексу $S_n$, т. е.,
\begin{equation}\label{eq1:8}
    \sum_{k=1}^n w_k(t)=1,\quad w_k(t)\geq0,\,k=1,\ldots,n.
\end{equation}
Аналогично интегральному симплексу определяются граничное множество $\bd S_n$ (существует по крайней мере один такой индекс $k$, что $w_k(t)=0$) и внутреннее множество $\Int S_n$ (для всех индексов $k=1,\ldots, n$, $w_k(t)$>0 для любого $t$). В силу структуры системы \eqref{eq1:8} множества $\bd S_n$ и $\Int S_n$ инвариантны.
\begin{remark}\label{remark:2} Любому элементу $\bs v(\bs x,t)\in S_n(\Omega_t)$ можно сопоставить элемент $\bs w(t)\in S_n$, положив $\bs w(t)=\overline{\bs v}(t)$, где, как и далее по тексту, черта обозначает среднее интегральное значение по $\Omega$:
$$
\overline{v}_k(t)=\int_\Omega v_k(\bs x,t)\D \bs x,\quad k=1,\ldots,n,\quad t\geq 0.
$$
\end{remark}

Введем следующее
\begin{definition}\label{def1:1}Система \eqref{eq1:1}--\eqref{eq1:3} называется биологически стабильной, если существуют такие $\varepsilon_0>0$ и $\delta_0>0$, что для всех компонент решения $\bs v(\bs x,t)\in S_n(\Omega_t)$ системы \eqref{eq1:1}--\eqref{eq1:3} выполняется
$$
\liminf_{t\to\infty}\|v_k(\bs x,t)\|\geq \varepsilon_0>0,\quad k=1,\ldots,n,
$$
при условии, что начальные условия \eqref{eq1:2} удовлетворяют условию
$$
\|\varphi_k(\bs x)\|\geq \delta_0>0.
$$
\end{definition}
Краткое обсуждение мотивации данного определения приведено в первом пункте. Здесь и далее $\|\cdot\|$ означает норму в $L^2(\Omega)$. Очевидным образом определение \ref{def1:1} переформулируется на случай нераспределенной задачи \eqref{eq1:7}.

Ряд необходимых и достаточных условий биологической стабильности системы \eqref{eq1:7} приведен в \cite{hofbauer1998ega}. Наиболее интересное для нас условие заключается в проверке неравенства
\begin{equation}\label{eq1:10}
    \DP{\bs{Aw}}{\bs p}-\DP{\bs{Aw}}{\bs w}>0
\end{equation}
для любых положений равновесия $\bs w\in\bd S_n$. Здесь $\bs p$ --- некоторый фиксированный элемент из $\Int S_n$, т. е.,
$$
\sum_{k=1}^n p_k=1,\quad p_k>0,\quad k=1,\ldots,n.
$$
По аналогии с \eqref{eq1:10} в \cite{novozhilov2013replicator} было получено достаточное условие биологической стабильности для распределенной системы \eqref{eq1:1}--\eqref{eq1:3}. Оно заключается в проверке условия \eqref{eq1:10} на любых элементах $\bs w\in \bd S_n$. При этом должно выполняться условие на величины параметров~$d_k$:
\begin{equation}\label{eq1:11}
    \lambda_1d_{\min}>\mu,
\end{equation}
где $d_{\min}=\min_k\{d_k\}$, $\mu$ --- спектральный радиус матрицы $\bs A$, $\lambda_1$ --- первое ненулевое собственное значение краевой задачи
\begin{equation}\label{eq1:12}
    -\Delta \psi (\bs x)=\lambda \psi(\bs x),\quad \bs x\in \Omega, \quad \left.\frac{\partial \psi}{\partial \nu}\right|_{\bs x\in \Gamma}=0.
\end{equation}

В \cite{novozhilov2013replicator} доказано, что если выполняется условие \eqref{eq1:11}, то все положения равновесия задачи \eqref{eq1:1}--\eqref{eq1:3} совпадают с положениями равновесия системы \eqref{eq1:7}, т. е. являются пространственно однородными. Таким образом, условия \eqref{eq1:10} и \eqref{eq1:11} обеспечивают биологическую стабильность системы только в том случае, когда положения равновесия этой системы пространственно однородны. Или, другими словами, доказанное в \cite{novozhilov2013replicator} достаточное условие биологической стабильности достаточно ограничительно тем, что противоречит в некотором смысле упомянутому в первом пункте неформальному наблюдению о том, что учет явной пространственной структуры часто влечет биологическую стабильность. В частности, если система \eqref{eq1:7} не является биологически стабильной, то и биологическую стабильность системы \eqref{eq1:1}--\eqref{eq1:3} установить не удастся, пользуясь доказанным условием.

В дальнейшем будем рассматривать случай, когда \eqref{eq1:11} не выполняется, и матрица взаимодействий $\bs A$ имеет по крайней мере одно положительное собственное значение.

\begin{definition} Будем говорить, что параметр $d_k$ системы \eqref{eq1:1}--\eqref{eq1:3} является $\mu$-резонансным, если существует собственное значение $\lambda_{s_k}$ задачи \eqref{eq1:12}, такое, что
\begin{equation}\label{eq1:14}
    d_k=\frac{\mu}{\lambda_{s_k}}\,,
\end{equation}
где $\mu$ --- положительные собственные значения матрицы $\bs A$.
\end{definition}

Остальная часть данной работы посвящена доказательству следующего факта: наличие $\mu$-резонансных параметров $d_k$ для некоторых $k$ может означать биологическую стабильность системы \eqref{eq1:1}--\eqref{eq1:3} тогда, когда соответствующая нераспределенная система \eqref{eq1:7} не является биологически стабильной.

\section{Существование пространственно неоднородных решений}В этом пункте мы докажем, что если существуют резонансные параметры $d_k$, удовлетворяющие \eqref{eq1:14}, то у системы \eqref{eq1:1}--\eqref{eq1:3} существуют пространственно неоднородные стационарные решения.

Для дальнейшего рассмотрим неравенство Пуанкаре в следующей форме (например,~\cite{rectoris1985}):
$$
R_0\int_{\Omega} u^2(\bs x)\D \bs x\leq\int_{\Omega}|\nabla u(\bs x)|^2\D \bs x+R_1\left(\int_\Omega u(\bs x)\D\bs x\right)^2,
$$
которое выполняется для всех $u(\bs x)\in W^{s,2}$, где $s=1$ или $s=2$. Здесь $|\bs v|^2=\DP{\bs v}{\bs v}$, $R_0,R_1$ --- положительные постоянные, которые зависят от геометрии области $\Omega$, но не зависят от $u(\bs x)$. В частном случае, когда
\begin{equation}\label{eq2:1}
    \int_{\Omega}u(\bs x)\D \bs x=0,
\end{equation}
неравенство принимает вид
\begin{equation}\label{eq2:2}
    R_0\int_\Omega u(\bs x)\D \bs x\leq \int_{\Omega}|\nabla u(\bs x)|^2\D\bs x.
\end{equation}
\begin{theorem}\label{theorem:1}Пусть $\bs{\hat{w}}=(\hat{w}_1,\ldots,\hat{w}_n)$ --- положение равновесия локальной репликаторной системы \eqref{eq1:7}, и пусть распределенная репликаторная система имеет $\mu$-резонансные параметры $d_k$, где $\mu$ --- положительные собственные значения матрицы $\bs A$. Тогда в окрестности положения равновесия $\bs{\hat{w}}$ существуют пространственно неоднородные решения системы \eqref{eq1:1}--\eqref{eq1:3}.
\end{theorem}
\begin{proof} Рассмотрим решения системы \eqref{eq1:1}, \eqref{eq1:2} с начальными условиями вида
\begin{equation}\label{eq2:3}
    v_k(\bs x,0)=\varphi_k(\bs x)+\varepsilon \xi_k(\bs x),\quad \varepsilon>0,
\end{equation}
где $\varphi_k(\bs x),\xi_k(\bs x)\in W^{s,2},\, s=1,2,\,k=1,\ldots,n$. Функции $\varphi_k(\bs x)$ и $\xi_k(\bs x)$ выбираем таким образом, чтобы
$$
\int_{\Omega}\varphi_k(\bs x)\D \bs x=\hat{w}_k,\quad \int_\Omega \xi_k(\bs x)\D \bs x=0,\quad k=1,\ldots,n.
$$
Будем искать решение системы \eqref{eq1:1}, \eqref{eq1:2} в виде
\begin{equation}\label{eq2:4}
    v_k(\bs x,t)=\hat{w}_k+\varepsilon \bigl(c_0^k(t)+V_k(\bs x,t)\bigr)+\varepsilon^2g_k(\bs x,t),\quad k=1,\ldots,n,
\end{equation}
где $g_k(\bs x,t)\in B(\Omega_t),\,g_k(\bs x,0)=0,\,k=1,\ldots,n$.

Возможность представления решения в форме \eqref{eq2:4} следует из полноты системы собственных функций $\psi_0(\bs x)=1,\,\{\psi_i(\bs x)\}_{i=1}^\infty$ задачи \eqref{eq1:12} в пространстве $W^{s,2},s=1,2$. В этом случае,
$$
\int_{\Omega} v_k(\bs x,t)\psi_0(\bs x)\D \bs x=\hat{w}_k+\varepsilon c_0^k(t),
$$
и
$$
V_k(\bs x,t)=\sum_{i=1}^\infty c_i^k \psi_i(\bs x),\quad k=1,\ldots,n.
$$
Так как для $\hat w_k$ выполняется \eqref{eq1:8}, то из \eqref{eq1:5} следует, что
\begin{equation}\label{eq2:5}
    \sum_{k=1}^n c_0^k(t)=0.
\end{equation}
Мы также использовали свойство ортонормальности системы $\{\psi_i(\bs x)\}_{i=0}^\infty$. Из этого свойства следует, что
\begin{equation}\label{eq2:6}
    \int_\Omega V_k(\bs x,t)\D\bs x=0.
\end{equation}
Подставим разложение \eqref{eq2:4} в систему \eqref{eq1:1} и проинтегрируем результат по области $\Omega$, учитывая при этом лишь члены порядка $\varepsilon$. Получим
\begin{equation}\label{eq2:6p}
    \dot c_0^k(t)=\hat{w}_k\Bigl(\bigl(\bs{Ac}_0(t)\bigr)_k-\DP{\bs{A}^\top\bs{\hat{w}}}{\bs c_0(t)}-\DP{\bs{A\hat{w}}}{\bs c_0(t)}\Bigr)+c_0^k(t)\Bigl(\bigl(\bs{A\hat{w}}\bigr)_k-\DP{\bs{A\hat{w}}}{\bs{\hat{w}}}\Bigr).
\end{equation}
Так как $\bs{\hat{w}}$ --- положение равновесия системы \eqref{eq1:7}, то
$$
\bigl(\bs{A\hat{w}}\bigr)_k=\DP{\bs{A\hat{w}}}{\bs{\hat{w}}}=\hat{f}^l.
$$
С другой стороны, в силу равенства \eqref{eq2:5},
$$
\DP{\bs{A\hat{w}}}{\bs c_0(t)}=\sum_{k=1}^n c_0^k(t)\bigl( \bs{A\hat{w}}\bigr)_k=\hat{f}^l\sum_{k=1}^nc_0^k(t)=0.
$$
Поэтому система \eqref{eq2:6p} принимает вид
\begin{equation}\label{eq2:7}
    \dot c_0^k(t)=\hat{w}_k\Bigl(\bigl(\bs{Ac}_0(t)\bigr)_k-\DP{\bs{A}^\top\bs{\hat{w}}}{\bs c_0(t)}\Bigr),\quad k=1,\ldots n.
\end{equation}
Матрица Якоби этой системы совпадает с матрицей Якоби системы \eqref{eq1:7}, вычисленной в $\bs{\hat{w}}$. Если $\bs{\hat{w}}$ --- асимптотически устойчива, то
$$
\lim_{t\to\infty} c_0^k(t)=0.
$$
В любом случае, начальные данные этой системы должны удовлетворять условию \eqref{eq2:5}. Следовательно, не умаляя общности, можно положить
$$
c_0^k(0)=0,\quad k=1,\ldots,n.
$$
Тогда система \eqref{eq2:7} имеет единственное решение $\bs c_0(t)\equiv 0$.

Если же при подстановке решений в виде \eqref{eq2:4} в систему \eqref{eq1:1} умножить полученный результат скалярно в $L^2(\Omega)$ на собственные функции $\psi_1(\bs x),\,\psi_2(\bs x),\ldots$ и учесть условия ортонормальности, то линейные по параметру $\varepsilon$ члены образуют систему уравнений относительно $\bs c_i(t)=\bigl(c_i^1(t),\ldots,c_i^n(t)\bigr)$
\begin{equation}\label{eq2:8}
    \bs{\dot c}_i(t)=\bs{\hat{W}}(\bs A-\lambda_i\bs D)\bs c_i(t),\quad i=1,2,\ldots,
\end{equation}
где $\bs{\hat{W}}=\diag(\hat{w}_1,\ldots,\hat{w}_n)$, $\bs D=\diag(d_1,\ldots,d_n)$. Из \eqref{eq2:3} следует, что начальные условия для этой системы определяются равенствами
$$
c_i^k(0)=\int_{\Omega} \xi_k(\bs x)\psi_i(\bs x)\D \bs x,\quad i=1,2,\ldots,\quad k=1,\ldots,n.
$$
По условию теоремы система \eqref{eq1:1} обладает $\mu$-резонансными параметрами $d_k$. Пусть, например, $d_1$ является $\mu$-резонансным, причем
$$
d_1=\frac{\mu}{\lambda_1}\,,
$$
где $\lambda_1$ --- первое ненулевое собственное значение задачи \eqref{eq1:12}. Тогда матрица задачи \eqref{eq2:8} имеет нулевое собственное значение, т.е., система \eqref{eq2:8} обладает нетривиальным решением $\bs{\hat c}_1=(\hat{c}_1^1,\ldots,\hat c_1^n)$, где все $c_1^k$ константы и не зависят от $t$.

Воспользуемся неравенством \eqref{eq2:2} в случае $u(\bs x)=V_1(\bs x,t)$, так как условие \eqref{eq2:1} выполнено в силу \eqref{eq2:6}:
$$
\|\nabla v_1(\bs x,t)\|=\varepsilon\|\nabla V_1(\bs x,t)\|+o(\varepsilon)\geq R_0\varepsilon \|V_1(\bs x,t)\|+o(\varepsilon)=R_0\varepsilon \sum_{k=1}^n(\hat c_1^k)^2+o(\varepsilon)\geq R_0\varepsilon\delta_0^2+o(\varepsilon),
$$
что означает, что
$$
\lim_{t\to\infty}\|\nabla v_1(\bs x,t)\|\geq \varepsilon R_0 \delta_0^2+o(\varepsilon)>0,
$$
т. е., что для любых $t>0$ существует пространственно неоднородное стационарное решение исходной системы.
\end{proof}

\begin{remark} Результат теоремы \ref{theorem:1} распространяется на случай, когда положение равновесия $\bs{\hat{w}}\in \bd S_n$. Пусть, например, $\hat{w}_k>0,\,k=1,\ldots,r$ и $\hat{w}_k=0,\,k=r+1,\ldots,n$. Обозначим через $\bs{\hat{A}}$ матрицу, которая получается из $\bs A$ вычеркиванием последних $n-r$ строк и столбцов. Если соответствующая репликаторная система \eqref{eq1:1} имеет $\hat{\mu}$-резонансные параметры, где $\hat{\mu}$ --- положительное собственное значение матрицы $\bs{\hat{A}}$, то в окрестности положения равновесия $\bs{\hat{w}}\in\bs S_n$ в пространстве $B(\Omega_t)$ существуют пространственно неоднородные решения системы \eqref{eq1:1}--\eqref{eq1:3}. Доказательство этого утверждения повторят аргументы доказательства теоремы \ref{theorem:1}.
\end{remark}

\begin{remark}\label{remark:3} Существенно подчеркнуть, что условия теоремы \ref{theorem:1} достаточны для существования пространственно неоднородных стационарных решений, но не необходимы, что демонстрируется в следующем примере.
\end{remark}
\begin{example}В качестве примера рассмотрим так называемую автокаталитическую распределенную репликаторную систему в случае $\Omega=[0,1]$:
\begin{equation}\label{eq2:9}
    \begin{split}
       \pdt u_1 & =u_1\left(au_1-f^s(\bs u)+d_1\frac{\partial^2 u_1}{\partial x^2}\right),\quad a>1, \\
        \pdt u_2&= u_2\left(u_2-f^s(\bs u)+d_2\frac{\partial^2 u_2}{\partial x^2}\right),\\
        f^s(\bs u)&=\int_0^1\left[au_1^2+u_2-d_1\left(\frac{\partial u_1}{\partial x}\right)^2-d_2\left(\frac{\partial u_2}{\partial x}\right)^2\right]\D x,\\
        \int_0^1\bigl(&u_1(x,t)+u_2(x,t)\bigr)\D x=1,\\
     \end{split}
\end{equation}
с граничными условиями
\begin{equation}\label{eq2:10}
    \frac{\partial u_i}{\partial x}(0,t)= \frac{\partial u_i}{\partial x}(1,t)=1,\quad i=1,2,
\end{equation}
и начальными условиями
\begin{equation}\label{eq2:11}
    u_1(x,0)=\begin{cases}1+\cos 2\pi x,&0\leq x<1/2,\\
    0,&1/2\leq x\leq 1,
    \end{cases}\qquad
        u_2(x,0)=\begin{cases}0,&0\leq x<1/2,\\
    1+\cos 2\pi x,&1/2\leq x\leq 1.
    \end{cases}
\end{equation}
Соответствующая нераспределенная репликаторная система вида \eqref{eq1:7} имеет единственное, в случае $a>1$, асимптотически устойчивое положение равновесия $\hat{w}_1=1,\,\hat{w}_2=0$. Предположим, что система \eqref{eq2:9}, \eqref{eq2:10} имеет резонансные параметры. Пусть, например,
\begin{equation}\label{eq2:12}
 d_1=\frac{a}{(2\pi)^2}\,,\quad d_2=\frac{1}{(2\pi)^2}\,,\quad a>1.
\end{equation}
Также, как и в случае теоремы \ref{theorem:1}, будем искать решение задачи \eqref{eq2:9}--\eqref{eq2:11} в виде
\begin{equation*}
    u_1(x,t)=\begin{cases}A(t)(1+\cos 2\pi x),&0\leq x<1/2,\\
    0,&1/2\leq x\leq 1,
    \end{cases}\quad
        u_2(x,t)=\begin{cases}0,&0\leq x<1/2,\\
    B(t)(1+\cos 2\pi x),&1/2\leq x\leq 1,
    \end{cases}
\end{equation*}
где $A(0)=B(0)=1$.

Из интегрального тождества в \eqref{eq2:9} следует, что
\begin{equation}\label{eq2:13}
    \frac 12 \bigl(A(t)+B(t)\bigr)=1,\quad t\geq 0.
\end{equation}
С учетом равенств \eqref{eq2:12} получим
$$
f^s(\bs u)=\frac 12 \bigl(aA^2(t)+B^2(t)\bigr).
$$
Если с помощью равенства \eqref{eq2:13} исключить функцию $B(t)$, то первое из уравнений \eqref{eq2:9}, после интегрирования по $x$, сведется к автономному уравнению первого порядка для неизвестной $A(t)$:
\begin{equation}\label{eq2:14}
    \dot A=-\frac 12 A\bigl((a+1)A^2-2(2+a)A+4\bigr).
\end{equation}
Правая часть этого уравнения имеет три корня $A_1=0,A_2=\frac{2}{a+1}\,,A_3=2$, которые являются положениями равновесия уравнения \eqref{eq2:14}. Так как $a>1$, то $A_2<A_3$. $A_2$ является аттрактором, с областью притяжения $0<A<2$. В частности, начальное значение $A(0)=1>\frac{2}{a+1}$ находится в этой области. Поэтому $\lim_{t\to\infty} A(t)=2$ и, соответственно, $B(t)\to 0$. Таким образом, решение $u_1(x,t)$ является пространственно неоднородным, причем
$$
\lim_{t\to\infty} \|\nabla u_1(x,t)\|=2\pi^2.
$$

Рассмотрим теперь случай, когда вместо равенств \eqref{eq2:12}, имеют место следующие неравенства:
$$
d_1<\frac{a}{(2\pi)^2}\,,\quad d_2<\frac{1}{(2\pi)^2}\,,
$$
и, кроме того,
$$
d_1=\frac{a}{(2\pi)^2}-\delta\,,\quad d_2<\frac{1}{(2\pi)^2}-\delta\,,\quad \delta>0.
$$
Тогда выражение для средней приспособленности примет вид
$$
f^s(\bs u)=a\left(\frac 12+\frac{\pi^2\delta}{a}\right)A^2(t)+\left(\frac 12+\pi^2\delta\right)B^2(t).
$$
В итоге, аналогично \eqref{eq2:14}, получим уравнение
$$
\dot A=-\frac 12 A\Bigl(\left(a+1+2\pi^2\delta\right)A^2-\left(2a+4+10\pi^2\delta\right)A+4-2\pi^2\delta\Bigr),
$$
для которого для достаточно малых $\delta$ справедливы те же рассуждения, как и в случае $\delta=0$. Это означает, что в данном примере возникают пространственно неоднородные решения как для резонансных величин параметров, так и для достаточно близких к ним.
\end{example}
\section{Достаточное условие биологической стабильности распределенной репликаторной системы}
Введем в рассмотрение следующую матрицу
$$
\bs H=R_0\bs D-\bs A,
$$
где, напомним, $R_0>0$ --- постоянная в неравенстве Пуанкаре \eqref{eq2:2}, и $\bs D=\diag(d_1,\ldots,d_n)$.
\begin{theorem}\label{theorem:2}Пусть система \eqref{eq1:1} с неотрицательной матрицей $\bs A$ имеет $\mu$-резонансные параметры $d_k$, удовлетворяющие равенству \eqref{eq1:14}. Пусть также выполняются условия
\begin{equation}\label{eq3:1}
 \DP{\bs{Aw} }{\bs p}-\DP{\bs{Aw}}{\bs w}\geq 0,\quad \bs p\in\Int S_n,
\end{equation}
для любых элементов $\bs w\in\bd S_n$, и
\begin{equation}\label{eq3:2}
\DP{\bs{Hz}}{\bs z}\geq 0,\quad \bs z=(z_1,\ldots,z_n)\in \R^n_+.
\end{equation}
Тогда распределенная репликаторная система \eqref{eq1:1}--\eqref{eq1:3} биологически стабильна.
\end{theorem}
\begin{proof}Рассмотрим функционал над решениями системы \eqref{eq1:1}--\eqref{eq1:3}:
$$
F(\bs v)=\exp\Bigl(\sum_{k=1}^np_k\overline{\log v_k(\bs x,t)}\Bigr),\quad \bs p\in\Int S_n,
$$
где
$$
\overline{\log v_k(\bs x,t)}=\int_\Omega \log v_k(\bs x,t)\D\bs x, \quad k=1,\ldots,n.
$$
Предположим, что начальные условия \eqref{eq1:3} ненулевые:
$$
v_k(\bs x,0)=\varphi_k(\bs x)>0,\quad \bs x\in\Omega,\quad k=1,\ldots,n.
$$
Тогда
\begin{equation}\label{eq3:3}
    F(\bs v)|_{t=0}=F_0>0.
\end{equation}
Если существует хотя бы одно решения задачи \eqref{eq1:1}--\eqref{eq1:3} такое, что $v_k(\bs x,t)\to 0$ при $t\to\infty$ для почти всех $\bs x\in\Omega$, то $F(\bs v)\to 0$ при $t\to \infty$. С другой стороны,
\begin{equation}\label{eq3:4}
    \frac{\D F}{\D t}(\bs v)=F(\bs v)\int_\Omega\bigl(\DP{\bs{Av}}{\bs p}-f^s(\bs v)\bigr)\D \bs x=F(\bs v)\int_\Omega G(\bs v)\D \bs x.
\end{equation}
Воспользуемся представлением решения системы \eqref{eq1:1}--\eqref{eq1:3} в форме
\begin{equation}\label{eq3:5}
    v_k(\bs x,t)=\overline{v}_k(t)+V_k(\bs x,t),\quad k=1,\ldots,n.
\end{equation}
Здесь
$$
\overline{v}_k(t)=\int_\Omega v_k(\bs x,t)\D\bs x,\quad V_k(\bs x,t)=\sum_{i=1}^\infty c_i^k(t)\psi_i(\bs x),
$$
где $\psi_i(\bs x)$ --- собственные функции задачи \eqref{eq1:12}. Подставляя \eqref{eq3:5} в правую часть равенства \eqref{eq3:4}, получим
$$
G(\bs v)=\DP{\bs{A\overline{v}}}{\bs p}-\DP{\bs{A\overline{v}}}{\bs{\overline{v}}}-\int_\Omega \DP{\bs{AV}}{\bs V}\D \bs x+\sum_{k=1}^nd_k\int_\Omega |\nabla V_k(\bs x,t)|^2\D\bs x,
$$
где также использовалось равенство \eqref{eq2:6}. Так как по предположению элементы матрицы $\bs A$ неотрицательны, то используя неравенство Коши--Буняковского, получим
$$
\int_\Omega \DP{\bs{AV}}{\bs V}\D \bs x\leq \sum_{k,l=1}^na_{kl}\|V_k\|\|V_l\|.
$$
С другой стороны, из неравенства Пуанкаре \eqref{eq2:2} следует, что
$$
\int_\Omega |\nabla V_k|^2\D\bs x\geq R_0\int_{\Omega}V_k^2\D\bs x.
$$
Отсюда,
$$
G(\bs v)\geq \DP{\bs{A\overline{v}}}{\bs p}-\DP{\bs{A\overline{v}}}{\bs{\overline{v}}}+\DP{\bs{Hz}}{\bs z},
$$
где $z_k=\|V_k\|,k=1,\ldots,n$.

Если выполняется условие \eqref{eq3:2}, то
\begin{equation}\label{eq3:9}
F(\bs v)\geq F_0\exp\Bigl(\int_0^t\bigl(\DP{\bs{A\overline{v}}}{\bs p}-\DP{\bs{A\overline{v}}}{\bs{\overline{v}}}\bigr)\D t \Bigr),\quad F_0>0.
\end{equation}

Используя замечание \ref{remark:2}, можно отождествить любую функцию $\bs v(\bs x,t)\in S_n(\Omega_t)$ с $\bs w(t)\in S_n$, положив $\bs w(t)=\bs{\overline{v}}(t)$. Если есть хотя бы одна компонента $v_k(\bs x,t)$ решения задачи \eqref{eq1:1}, такая, что $v_k(\bs x,t)\to 0$ при $t\to\infty$, то $\overline{v}_k(t)\to0$ при $t\to\infty$, следовательно, $F(\bs v)\to 0$. С другой стороны, из неравенства \eqref{eq3:9} и условия \eqref{eq3:1} следует, что
$$
F(\bs v)\geq F_0>0,\quad t\geq 0,
$$
следовательно, $v_k(\bs x,t)$ не стремятся к нулю, и, следовательно, система биологически стабильна.
\end{proof}
\begin{corollary}\label{corr:1} Пусть
\begin{equation}\label{eq3:10}
    \min_{\bs w\in S_n}\bigl\{\DP{\bs{Aw}}{\bs p}-\DP{\bs{Aw}}{\bs w}\bigr\}=-m<0,\quad \bs p\in\Int S_n,
\end{equation}
и выполняется условие положительной определенности формы
\begin{equation}\label{eq3:11}
    \DP{\bs{Hz}}{\bs z}\geq \gamma |\bs z|^2,\quad \gamma>0,\quad \bs z\in\R^n_+,
\end{equation}
где $|\bs z|=(z_1^2+\ldots+z_n^2)^{1/2}$.

Тогда, при наличии $\mu$-резонансных параметров $d_k$, и если существует хотя бы одна компонента $V_k(\bs x,t)$ решения
$$
v_k(\bs x,t)=\overline{v}_k(t)+V_k(\bs x,t),\quad k=1,\ldots,n,
$$
которая удовлетворяет условию
\begin{equation}\label{eq3:12}
    \lim_{t\to\infty} \|V_k\|^2\geq \frac{m}{\gamma}\,,
\end{equation}
то система \eqref{eq1:1}--\eqref{eq1:3} биологически стабильна.
\end{corollary}
\begin{proof}Прежде всего отметим, что в рассматриваемом случае достаточные условия биологической стабильности нераспределенной репликаторной системы \eqref{eq1:7} могут не выполняться. Таким образом, речь идет о варианте, когда наличие пространственной структуры влечет биологическую стабильность системы.  Отметим также, что теорема \ref{theorem:1} гарантирует существование пространственно неоднородных решений системы \eqref{eq1:1} в окрестности положений равновесия системы \eqref{eq1:7} в пространстве $B(\Omega_t)$.

Остальное доказательство повторяет аргументы доказательства теоремы \ref{theorem:2}. В рассматриваемом случае в силу \eqref{eq3:10} и \eqref{eq3:11} справедливо неравенство
$$
G(\bs v)\geq -m+\gamma |\bs z|^2,\quad z_k=\|V_k\|,\quad k=1,\ldots,n,
$$
и если выполняется условие \eqref{eq3:12}, то
$$F(\bs v)\geq F_0>0,\quad t\geq 0.
$$
\end{proof}
\section{Примеры}
В этом пункте мы иллюстрируем численно общие утверждения теорем \ref{theorem:1}, \ref{theorem:2} и следствия \ref{corr:1} примерами.
\begin{example}Рассмотрим распределенную репликаторную систему, заданную на $\Omega=[0,1]$:
\begin{equation}\label{eq4:1}
    \begin{split}
      \pdt v_1 & =v_1\left(av_1+kv_2-f^s(\bs v)+d_1\frac{\partial v_1}{\partial x^2}\right), \\
      \pdt v_2 & =v_2\left(kv_1-f^s(\bs v)+d_1\frac{\partial v_2}{\partial x^2}\right),
    \end{split}
\end{equation}
с начальными и краевыми условиями
\begin{equation}\label{eq4:2}
    v_1(x,0)=\varphi_1(x),\quad v_2(x,0)=\varphi_2(x),\quad \frac{\partial v_i}{\partial x}(0,t)=\frac{\partial v_i}{\partial x}(1,t)=0,\quad i=1,2.
\end{equation}
В силу инвариантности интегрального симплекса $S_2(\Omega_t)$ мы также имеем
\begin{equation}\label{eq4:3}
    \begin{split}
      \int_0^1\bigl(&v_1(x,t)+v_2(x,t)\bigr)\D x =1 \\
     f^s(\bs v)&=\int_0^1\left(2kv_1v_2+av_2^2-d_1\left(\frac{\partial v_1}{\partial x}\right)^2-d_2\left(\frac{\partial v_2}{\partial x}\right)^2\right)\D x.
    \end{split}
\end{equation}
Соответствующая нераспределенная репликаторная система
\begin{equation}\label{eq4:4}
    \begin{split}
      \dot w_1&=w_1\bigl(aw_1+kw_2-f^l(\bs w)\bigr),\\
       \dot w_2&=w_2\bigl(kw_1-f^l(\bs w)\bigr),
    \end{split}
\end{equation}
с условиями
\begin{equation}\label{eq4:5}
    w_1(0)=w_1^0,\quad w_2(0)=w_2^0,\quad w_1(t)+w_2(t)=1, \quad f^l(\bs w)=2kw_1w_2+aw_1^2.
\end{equation}
Если $a>k$, то система \eqref{eq4:4}, \eqref{eq4:5} имеет единственное асимптотически устойчивое положение равновесия $\bs{\hat{w}}=(1,0)$ и, следовательно, не является биологически стабильной.

В случае $a=k$ рассмотрим функцию $W=\log w_1(t)$. Тогда
$$
\dot W=a(1-w_1^2-2w_1w_2)=a(1-w_1)^2\geq 0.
$$
Следовательно и в этом случае положение равновесия $\bs{\hat{w}}$ является омега предельным множеством.

При резонансных параметрах $d_k$ у распределенной системы \eqref{eq4:1} можно ожидать появления пространственно неоднородных стационарных положений равновесия и, как следствие, биологической стабильности системы. Действительно, численные расчеты показывают, что, взяв матрицу взаимодействий в форме
\begin{equation}\label{eq:A:1}
\bs A=\begin{bmatrix}
        1.1 & 1 \\
        1 & 0 \\
      \end{bmatrix},
\end{equation}
и параметры
$$
\bs d=\frac{1}{\pi^2}(1.1,0.5),
$$
то обе популяция взаимодействующих макромолекул сосуществуют (см. рис. \ref{fig:n1} и \ref{fig:n2}).
\begin{figure}[!th]
\centering
\includegraphics[width=0.7\textwidth]{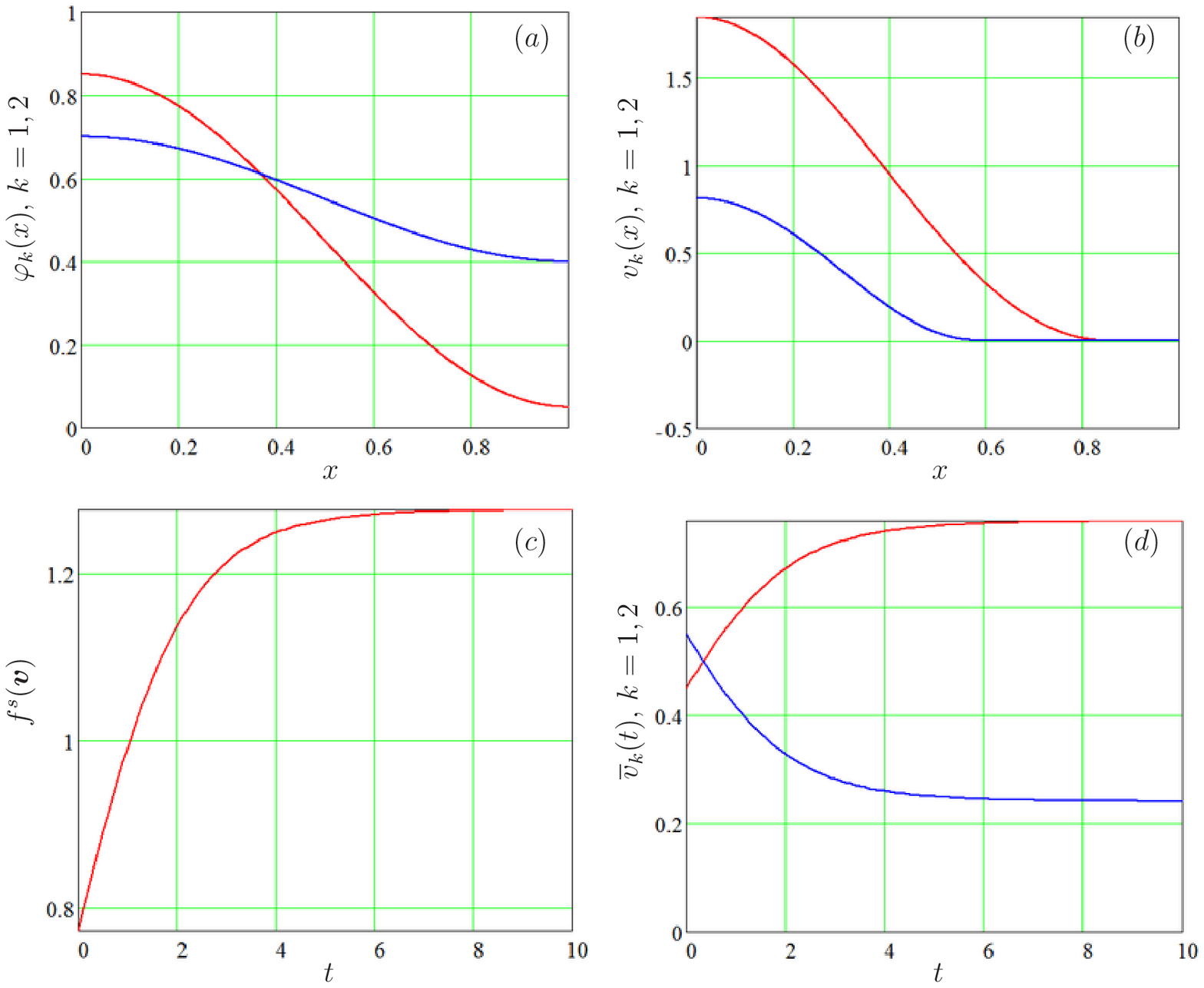}
\caption{Решения задачи \eqref{eq1:1} с матрицей взаимодействий \eqref{eq:A:1}. $(a)$ Начальные условия. $(b)$ Решения в момент времени $t=10$. $(c)$ Зависимость средней приспособленности от времени. $(d)$ Зависимость от времени средних интегральных выражений $\overline{v}_k(t)=\int_\Omega v_k(x,t)\D x$. См. также рис. \ref{fig:n2}, где показаны решения задачи \eqref{eq1:1} на плоскости $(x,t)$ }\label{fig:n1}
\end{figure}

\begin{figure}[!bh]
\centering
\includegraphics[width=0.6\textwidth]{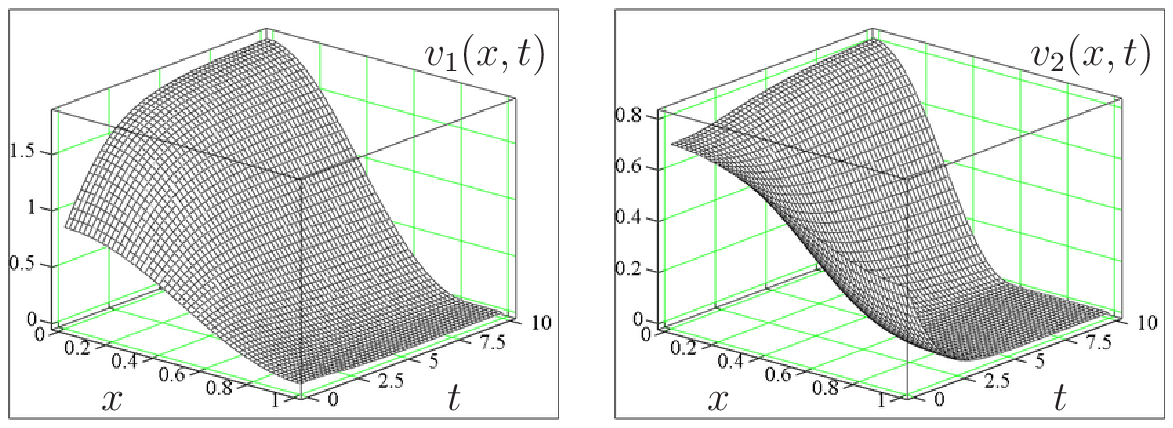}
\caption{Решения задачи \eqref{eq1:1} с матрицей взаимодействий \eqref{eq:A:1}}\label{fig:n2}
\end{figure}

Характерной особенностью полученных стационарных пространственно-неоднородных решений системы \eqref{eq4:1} (см. ) является тот факт, что носитель этих решений не совпадает с интервалом $(0,1)$. Это явление прослеживается и в других численных примерах, например см. рис. \ref{fig:n3} и \ref{fig:n4}, где расчет выполнялся для матрицы взаимодействий
\begin{equation}\label{eq:A:2}
\bs A=\begin{bmatrix}
        1.1 & 0 & 1 \\
        1 & 0 & 0 \\
        0 & 1 & 0 \\
      \end{bmatrix},
\end{equation}
и параметров
$$
\bs d=\frac{1}{\pi^2}(1.1,0.3,0.1).
$$
\begin{figure}[!t]
\centering
\includegraphics[width=0.7\textwidth]{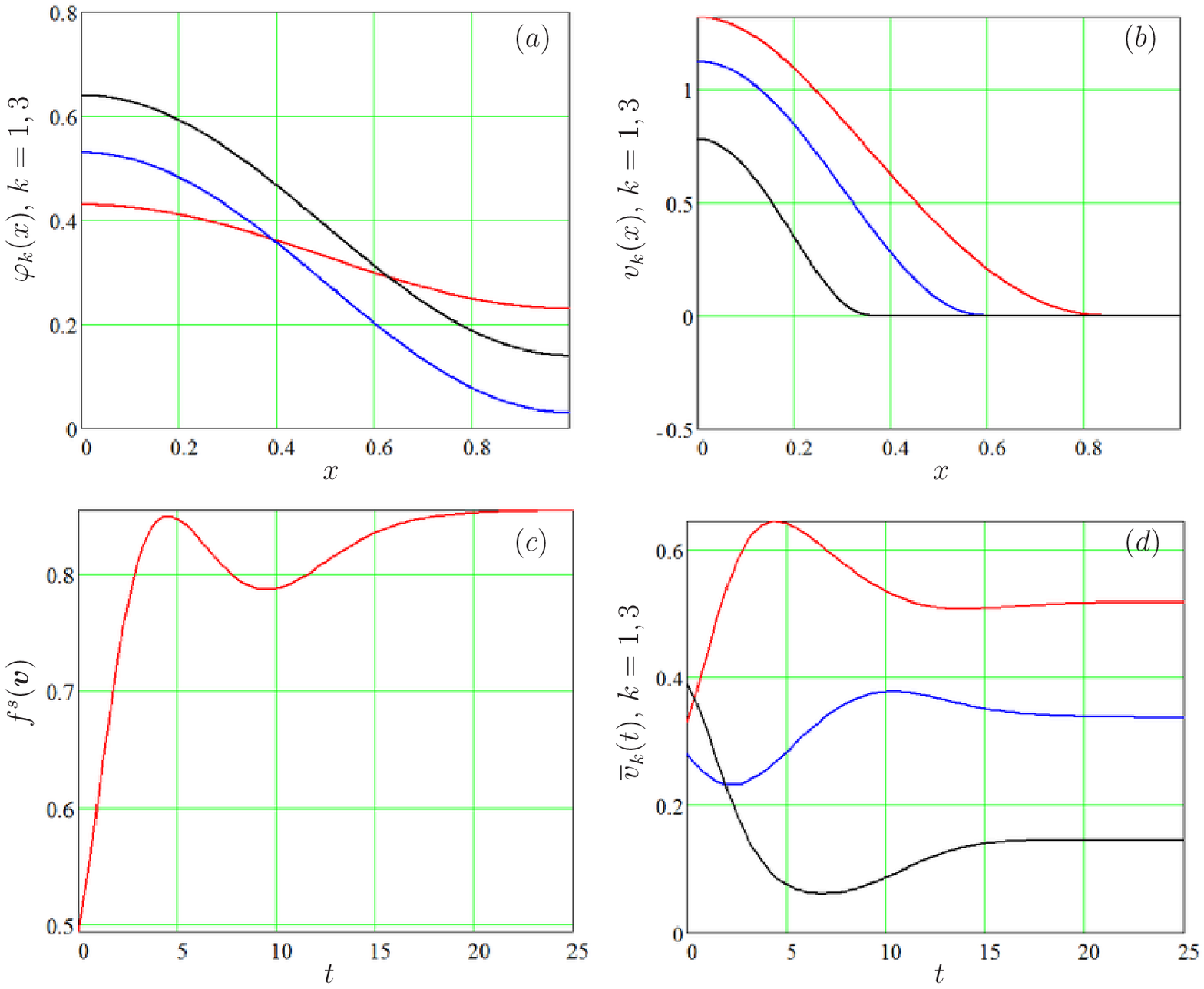}
\caption{Решения задачи \eqref{eq1:1} с матрицей взаимодействий \eqref{eq:A:2}. $(a)$ Начальные условия. $(b)$ Решения в момент времени $t=25$. $(c)$ Зависимость средней приспособленности от времени. $(d)$ Зависимость от времени средних интегральных выражений $\overline{v}_k(t)=\int_\Omega v_k(x,t)\D x$. См. также рис. \ref{fig:n4}, где показаны решения задачи \eqref{eq1:1} на плоскости $(x,t)$ }\label{fig:n3}
\end{figure}
\begin{figure}[!bh]
\centering
\includegraphics[width=0.9\textwidth]{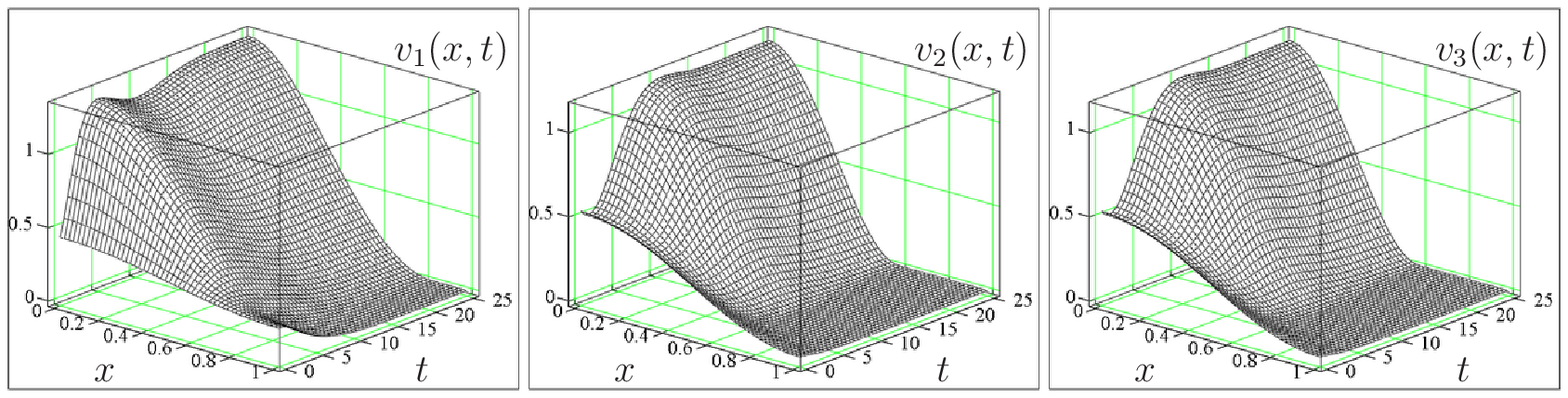}
\caption{Решения задачи \eqref{eq1:1} с матрицей взаимодействий \eqref{eq:A:2}}\label{fig:n4}
\end{figure}
\end{example}

\begin{example} В качестве второго примера рассмотрим систему, которая основана на \textit{in vitro} экспериментах с РНК молекулами \cite{vaidya2012spontaneous}. Матрица взаимодействий имеет вид
\begin{equation}\label{eq5:1}
    \bs A=\left[
\begin{array}{cccccc}
 0 & 0 & \alpha  & 0 & 0 & \gamma  \\
 \alpha  & 0 & 0 & 0 & \gamma  & 0 \\
 0 & \alpha  & 0 & \gamma  & 0 & 0 \\
 \gamma  & 0 & 0 & \beta  & 0 & 0 \\
 0 & 0 & \gamma  & 0 & \beta  & 0 \\
 0 & \gamma  & 0 & 0 & 0 & \beta  \\
\end{array}
\right].
\end{equation}
Если взять величины параметров
$$
\alpha=1.75,\quad \beta=0.7,\quad \gamma=2.0,
$$
и
$$
\bs d=(0.04,0.05,0.04,0.05,0.04,0.05),
$$
то, в отличие от локальной системы, в которой концентрации трех из шести макромолекул всегда стремятся к нулю, распределенная система является биологически стабильной, при этом как показывают численные эксперименты, решения не выходят на стационарное решения, а являются периодическими по времени (см. рис. \ref{fig:2}).
\begin{figure}[!bh]
\centering
\includegraphics[width=0.4\textwidth]{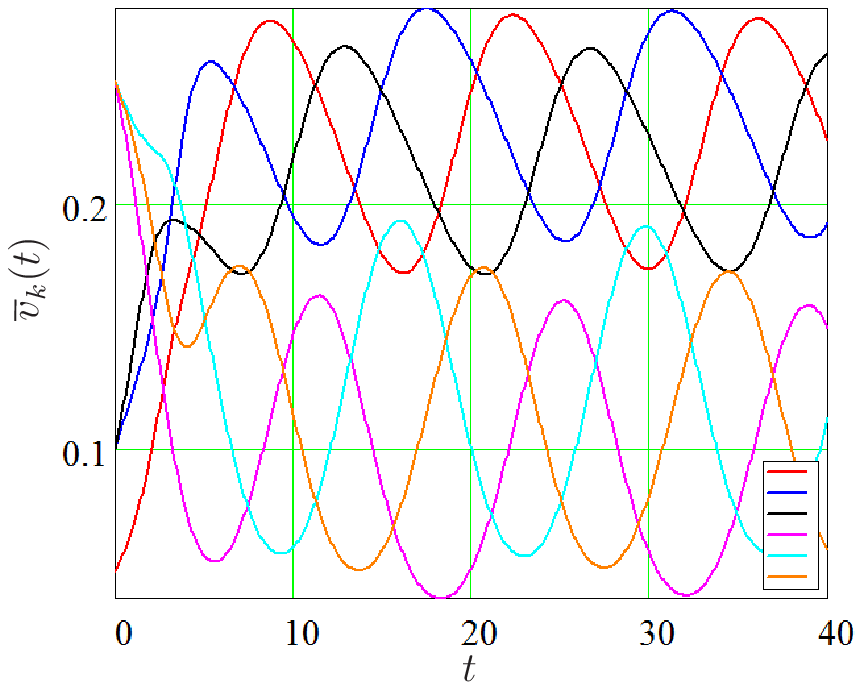}
\caption{Решения задачи \eqref{eq1:1} с матрицей взаимодействий \eqref{eq5:1}. Показаны интегральные средние значения переменных в зависимости от времени }\label{fig:2}
\end{figure}

Подробнее этот пример рассматривается в \cite{novozhilov2013replicator}.
\end{example}

\paragraph{Благодарности:} Исследования поддержаны Российским Фондом Фундаментальных Исследований (РФФИ) грант № 10-01-00374 и совместным грантом РФФИ и Тайваньского Национального Фонда, грант № 12-01-92004HHC-a. АСН поддержан грантом Национального Научного Фонда США (NSF USA) № EPS-0814442.

\end{document}